\documentclass[a4paper,11pt]{article}

\usepackage[top=18truemm,bottom=25truemm,left=25truemm,right=25truemm]{geometry} 
\usepackage{indentfirst} 

\makeatletter 
\def\section{\@startsection {section}{1}{\z@}{-3.5ex plus -1ex minus -.2ex}{2.3 ex plus .2ex}{\LARGE\bf}}
\makeatother
\usepackage{secdot} 

\makeatletter 
\def\subsection{\@startsection {subsection}{1}{\z@}{-3.5ex plus -1ex minus -.2ex}{2.3 ex plus .2ex}{\Large\it}}
\makeatother

\usepackage{amsmath, amssymb} 

\usepackage[dvipdfmx]{graphicx}
\usepackage{float} 
\graphicspath{{./}} 
\usepackage[labelsep=period,figurename=Fig. ,tablename=Table]{caption} 
\usepackage{subcaption} 
\usepackage{booktabs}
\usepackage{multirow} 
\usepackage{tabularx}
\newcolumntype{C}{>{\centering\arraybackslash}X}

\usepackage{cite}
\usepackage{url}

\usepackage{comment} 
\usepackage{color}
\usepackage{ascmac}
\usepackage{fancybox}

\def\vector#1{\mbox{\boldmath $#1$}}

\title{Statistical properties for directional alignment and chasing of players in football games}

\usepackage{authblk}
\author[1]{Takuma Narizuka\thanks{{\it E-mail address}: pararel@gmail.com (T. Narizuka).\\
}}
\author[1]{Yoshihiro Yamazaki}
\affil[1]{Department of Physics, School of Advanced Science and Engineering, Waseda University, Shinjuku, Tokyo 169-8555, Japan}
\date{}

\begin{document}
	\maketitle
\begin{abstract}
Focusing on motion of two interacting players in football games, two velocity vectors for the pair of one player and the nearest opponent player exhibit strong alignment.
Especially, we find that there exists a characteristic interpersonal distance $ r\simeq 500 $ cm below which the circular variance for their alignment decreases rapidly.
By introducing the order parameter $ \phi(t) $ in order to measure degree of alignment of players' velocity vectors, we also find that the angle distribution between the nearest players' velocity vectors becomes wrapped Cauchy ($ \phi \lesssim 0.7 $) and the mixture of von Mises and wrapped Cauchy distributions ($ \phi \gtrsim 0.7 $), respectively.
To understand these findings, we construct a simple model for the motion of the two interacting players with the following rules: chasing between the players and the reset of the chasing.
We numerically show that our model successfully reproduce the results obtained from the actual data.
Moreover, from the numerical study, we find that there is another characteristic distance $ r\simeq 1000 $ cm below which player's chasing starts.
\end{abstract}

\baselineskip 20pt
\section{Introduction}
Competition is ubiquitously observed in a wide range of physical, biological, and social systems, including predator-prey interactions \cite{Murray2002}, geometric frustration \cite{Moessner2006}, chemical reaction \cite{Winfree1974}, hierarchy formation \cite{bonabeau1995}，resource allocation \cite{Chakraborti2015}，and sports \cite{Ben-Naim2012}．
For studying human behaviours, competitive sports seem to be appropriate systems in that they realize the human competition in a particular place and time interval.
Generally, competitive sports have the following essential features.
First, all players must follow unified rules restricting their motions.
When each player and each team attempt to win the game under these rules, various human behaviours emerge: scoring events, ball-passing, marking an opponent player, keeping formation, for example.
Second, interactions among players include uncertainty, and outcomes of game are unpredictable. 
The players' motions with uncertainty make it possible to describe the above behaviours by stochastic processes on different time scales, from seconds and minutes to hours.
Our main object is to clarify their statistical properties.

For analysing these behaviours in competitive sports, the viewpoint of statistical physics seems to be available.
One of the successful examples is application of a random walk to the scoring events.
So far, it has been found that scoring events of some sports are well described by a Poisson process \cite{Thomas2007, Heuer2010, Gabel2012, Merritt2014}.
Several studies have revealed existence of the arcsin law \cite{Clauset2015} and anomalous diffusion \cite{Ribeiro2012, DaSilva2013, Kiley2015} of the time series of scoring events, the Zipf--Mandelbrot law of the goal distribution \cite{Malacarne2000}, and so on.
Meanwhile, ball motion on the field is essential in football.
Actually, complex network analysis of the ball-passing \cite{Duch2010, Pena2012, Narizuka2014, Narizuka2015}, fractal analysis of the ball motion \cite{Kijima2014}, and statistical analysis of the ball-possession time \cite{Mendes2007} have been done.

Interaction among players is another important topic in studying of competitive sports.
In particular, cooperative behaviours arising among interacting players are of main interest.
Previous studies have focused on the dynamics of the relative phase so as to measure the cooperation of the players \cite{Lamb2014, Davids2014}.
It has been reported that in-phase and anti-phase cooperation patterns emerge in player motions of various competitive sports \cite{Palut2005, McGarry2006, Bourbousson2010a, Siegle2013}.
In the present study, we give quantitative characterization and theoretical consideration for the player interactions in football games.

\section{Method}
\begin{figure}[htbp]
	\centering
	\includegraphics[width=14cm]{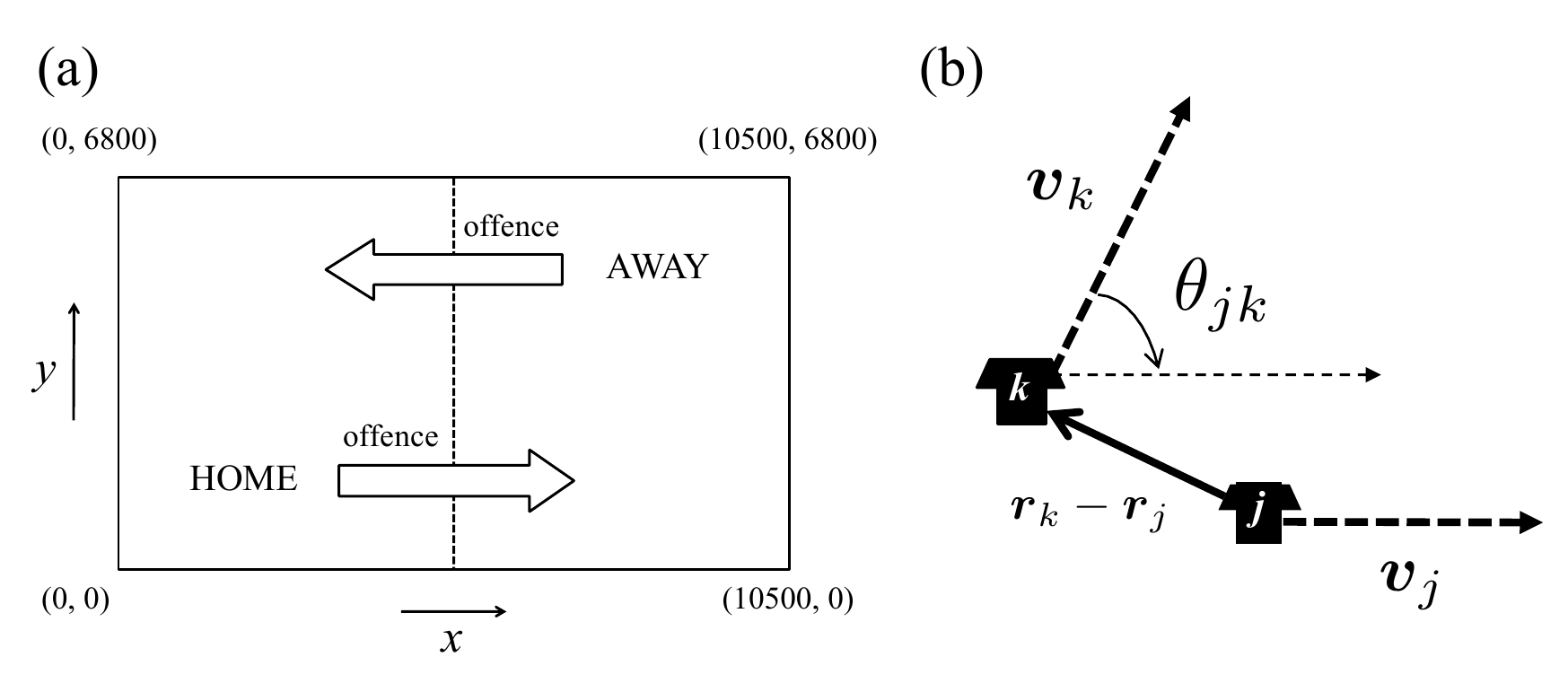}
	\vspace{-0.3cm}
	\caption{(a) The coordinate system for football field. The goal direction corresponds to $ x $ axis, and the vertical direction is $ y $ axis.
A centimetre unit is used, and the domain of each axis is $ 0 \leq x \leq 10500 $ and $ 0 \leq y \leq 6800 $, respectively.
The direction of offence for home team is rightward.
(b) Definition of angle $ \theta_{jk}(t) $ between two velocity vectors, $ \vector{v}_{j} $ and $ \vector{v}_{k} $.
}
	\label{fig:field_angle}
\end{figure}

The player tracking data of 11 matches in the Japan Professional Football League 2015 and 2016 were provided by DataStadium Inc., Japan.
Each game is distinguished by labels from ``G1'' to ``G11'' (see table \ref{tb:game}).
All player positions are recorded every $ \Delta t = 0.04 $ seconds in the coordinate system illustrated in fig. \ref{fig:field_angle}(a).
The position of $ j $-th player at time $ t $ is shown as $ \vector{r}_{j}(t) = [x_{j}(t), y_{j}(t)] $, and the interpersonal distance $ r_{jk}(t) $ is defined as $ r_{jk}(t)=|\vector{r}_{k}(t) - \vector{r}_{j}(t)| $.

\begin{figure}[htbp]
	\centering
	\includegraphics[width=14cm]{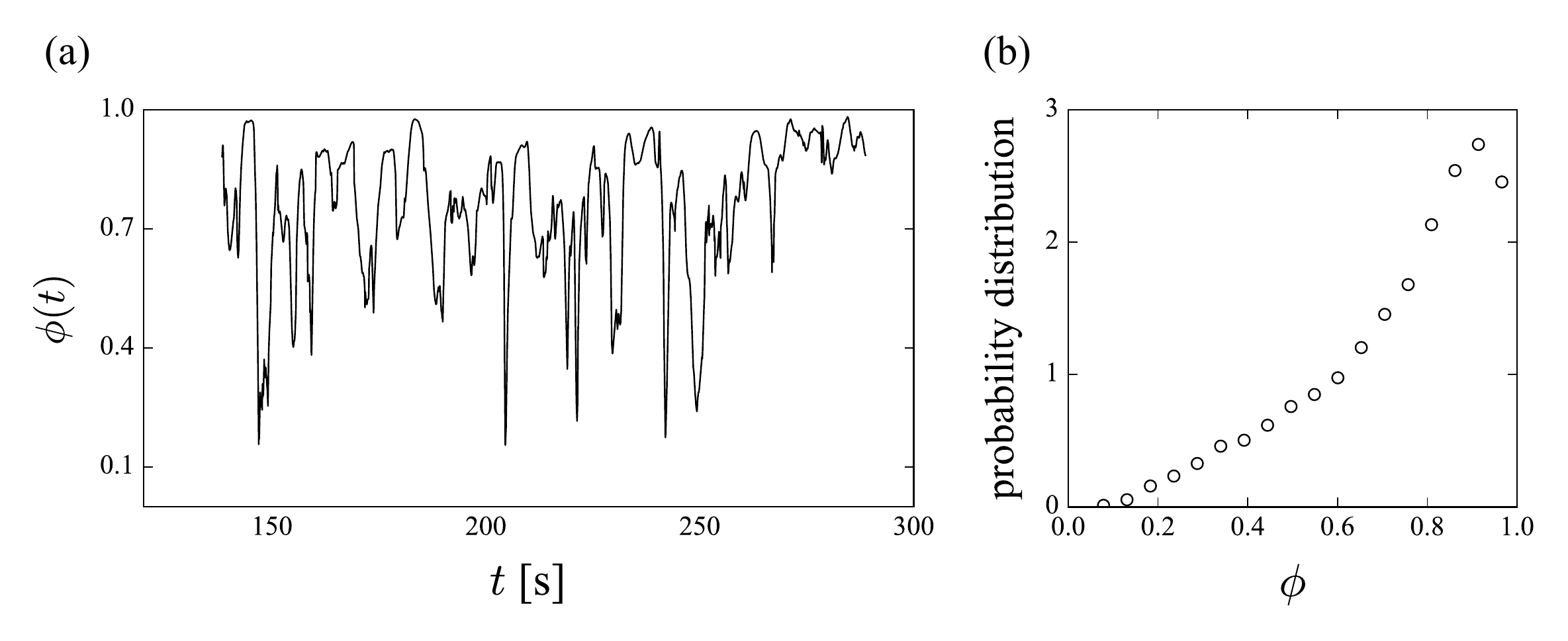}
	\vspace{-0.3cm}
	\caption{(a) A typical time sequence and (b) probability distribution of $ \phi(t) $ obtained from G10.}
	\label{fig:op-t_Dop}
\end{figure}

The velocity vector of the $ j $-th player is calculated as
\begin{align}
	\vector{v}_{j}(t)
	&= \frac{\vector{r}_{j}(t + n\Delta t) - \vector{r}_{j}(t)}{n\Delta t},
\end{align}
where $ \Delta t = 0.04 $ s and we set $ n = 25 $.
The alignment of moving directions among players is measured by the following two ways.
(i) For the pairs of two players $ j $ and $ k $, we focus on the angle $ \theta_{jk}(t) $ between their velocity vectors (see fig. \ref{fig:field_angle}(b)).
In the next section, we calculate $ \theta_{jk}(t) $ for all pairs of $ j $-th player and the $ k $-th nearest opponent player to the $ j $-th player every 0.2 seconds.
Statistics of $ \theta $ are characterized on a circle: the angle distribution $ f(\theta) $ and its circular variance $ V_{\theta} $.
Here, $ V_{\theta} $ is calculated from angle data $ \{\theta_{i}; i=1,\ldots, M \} $ as follows \cite{Mardia}:
\begin{align}
	V_{\theta} &= 1 - \bar{R},
\end{align}
where $ \bar{R} $ is the mean resultant length given by
\begin{align}
	\bar{R} 
	&= \left|\frac{1}{M} \sum_{i=1}^{M} (\cos\theta_{i}, \sin\theta_{i}) \right|.
\end{align}
Here, $ M $ is the length of the angle data.
The domain of circular variance is $ 0 \leq V_{\theta} \leq 1 $.
When $ V_{\theta} $ becomes smaller, the alignment is stronger.
(ii) For the measure of the directional alignment of each team, the order parameter $ \phi(t) $ is introduced as
\begin{align}
	\phi(t) &= \left|\frac{1}{N} \sum_{j=1}^{N} \frac{\vector{v}_{j}(t)}{|\vector{v}_{j}(t)|}\right|,
\end{align}
where $ N = 20 $ is the sum of the number of players except for the goal keepers in each team.
Note that $ 0 \leq \phi(t) \leq 1$, and $ \phi(t) $ approaches 1 as all players move in the same direction.
A typical time sequence and probability distribution of $ \phi(t) $, fig. \ref{fig:op-t_Dop}, shows that large and small $ \phi(t) $ appear repeatedly.

In the following analysis, $ V_{\theta} $ and $ f(\theta) $ are calculated from the angle data $ \{\theta_{jk}(t)\} $ where $ j,\ k $ and $ t $ satisfy $ \phi \leq \phi(t) < \phi + \Delta\phi $ or $ r \leq r_{jk}(t) < r + \Delta r $.

\begin{table}[htbp]
  \centering
  \caption{Actual game data used for the analysis.}
  \vspace{-0.2cm}
    \begin{tabular}{cccc}
    \toprule
 Label& Game (Home -- Away)   & Score & Date  \\ \toprule
 G1   & Iwata -- Nagoya       & 0--1  & 2016.02.27  \\
 G2   & Hiroshima -- Kawasaki & 0--1  & 2016.02.27  \\
 G3   & Tosu -- Fukuoka       & 2--1  & 2016.02.27  \\
 G4   & Kashiwa -- Urawa      & 1--2  & 2016.02.27  \\
 G5   & Shonan -- Niigata     & 1--2  & 2016.02.27  \\ 
 G6   & Kobe -- Kofu          & 0--2  & 2016.02.27  \\
 G7   & Yokohama -- Sendai    & 0--1  & 2016.02.27  \\ 
 G8   & Tokyo -- Omiya        & 0--1  & 2016.02.27  \\
 G9   & Osaka -- Kashima      & 0--1  & 2016.02.28  \\
 G10  & Kashima -- Shonan     & 1--2  & 2015.03.14  \\ 
 G11  & Matsumoto -- Shonan   & 2--3  & 2015.06.27  \\ 
    \bottomrule 
    \end{tabular}
  \label{tb:game}
\end{table}
%

\section{Data analysis}
\begin{figure}[htbp]
	\centering
	\includegraphics[width=10cm]{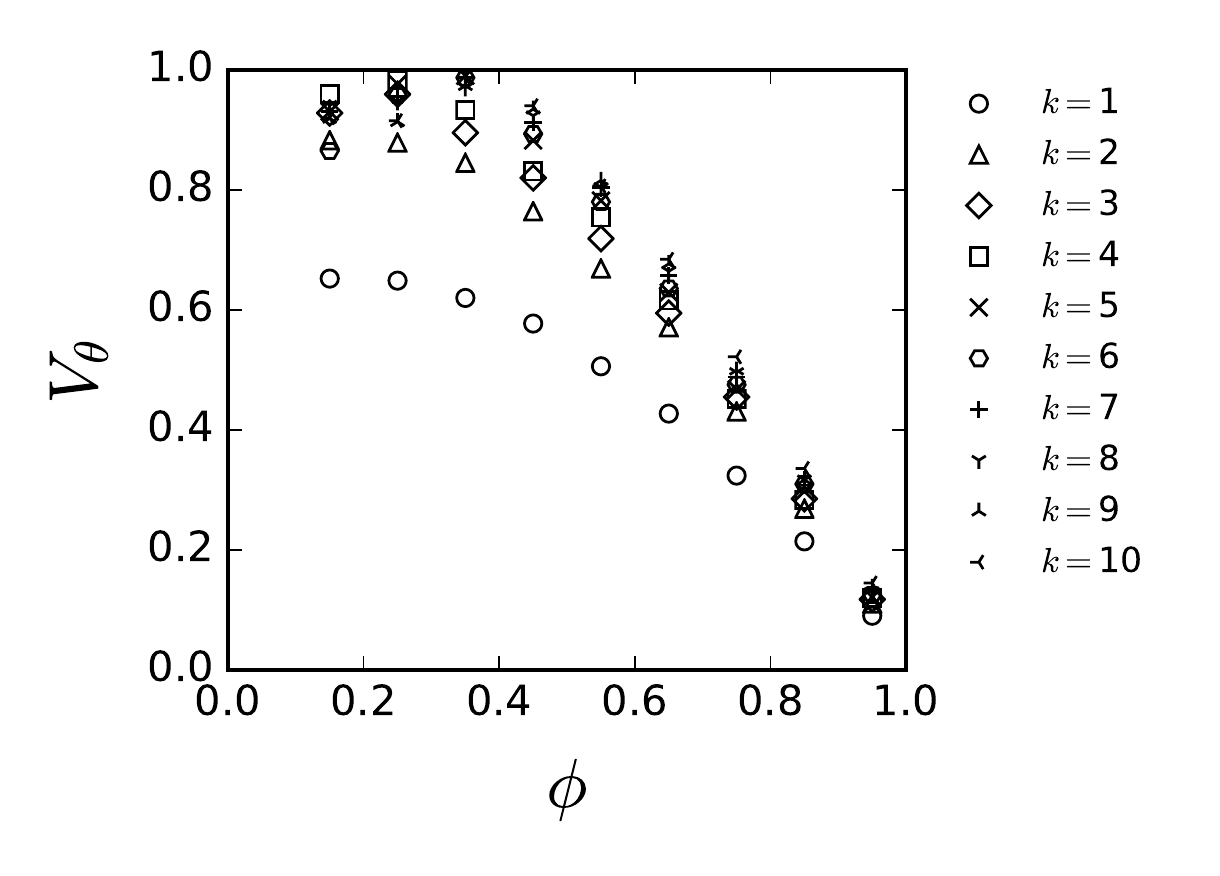}
	\vspace{-0.3cm}
	\caption{Circular variance $ V_{\theta} $ as a function of $ \phi $ obtained from the home team in G10.}
	\label{fig:op-agsd}
\end{figure}

For the pair of $ j $-th player and the $ k $-th nearest opponent player to the $ j $-th player, we calculate the angle $ \theta_{jk}(t) $ for all pairs of $ j $ and $ k $ every 0.2 seconds.
In the present paper, results for one team or game are mainly shown, though we have confirmed that almost the same results are obtained from the others.
Figure \ref{fig:op-agsd} shows the circular variance $ V_{\theta} $ as a function of $ \phi $ for different $ k $.
We find that when $ \phi $ is close to 1, $ V_{\theta} $ becomes low independently of $ k $.
And a remarkable difference between $ k=1 $ and $ k \geq 2 $ emerges with decreasing of $ \phi $; the plot for $ k=1 $ keeps low circular variance compared with those for $ k\geq2 $.

Next, we examine angle distribution $ f(\theta; \phi) $ for the nearest pair ($ k=1 $).
In order to characterize the obtained distributions, we introduce well-known common distribution functions on the circle \cite{Mardia}.
The first one is the von Mises distribution:
\begin{align}
	\mathrm{VM}(\theta, \kappa)
	&= \frac{1}{2\pi I_{0}(\kappa)} \mathrm{e}^{\kappa \cos\theta},
\end{align}
where $ I_{0}(\kappa) $ denotes the modified Bessel function of the first kind and order 0, and $ \kappa $ is the concentration parameter.
VM can be derived from the diffusion process with a drift on the circle.
Another one is the wrapped Cauchy distribution:
\begin{align}
	\mathrm{WC}(\theta, \rho) 
	&= \frac{1}{2\pi}\frac{1-\rho^{2}}{1+\rho^{2} -2\rho\cos\theta },
\end{align}
where $ \rho $ is the parameter which equals to the mean resultant length.
The feature of WC is that it has a sharp peak and fat tails compared with VM.
It is often used to generate the turning angle in the simulation of animal movement \cite{Wu2000, Bartumeus2005, Bartumeus2008, Codling2008, Codling2010}.

\begin{figure*}[htbp]
	\centering
	\includegraphics[width=16cm]{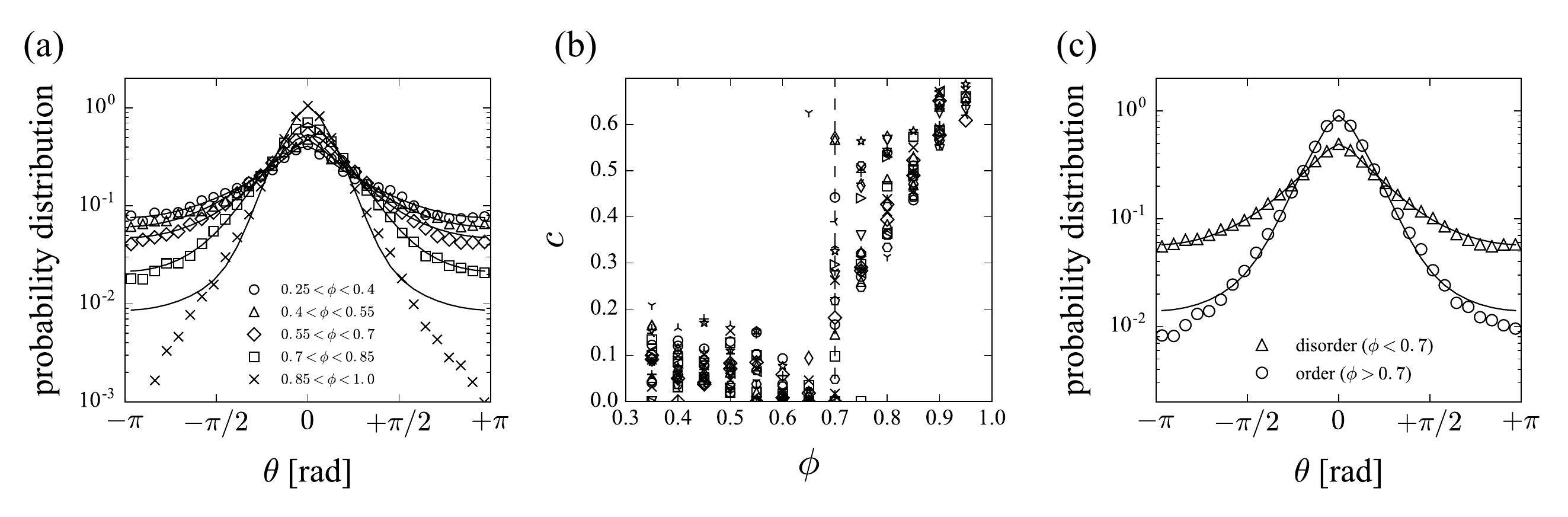}
	\vspace{-0.3cm}
	\caption{(a) Angle distributions for different values of $ \phi $ obtained from the home team in G10. The solid curves show the distribution function of eq. \eqref{eq:mvw}. (b) Mixing weight $ c $ as a function of $ \phi $ for all teams. There is a point ($ \phi \simeq 0.7 $) at which the value of $ c $ changes drastically. (c) Angle distributions for order and disorder phases obtained from the home team in G10. The solid curves show the distribution function of eq. \eqref{eq:mvw}.}
	\label{fig:Dag-op_op-c_Dag-07}
\end{figure*}

As shown in fig. \ref{fig:Dag-op_op-c_Dag-07}(a), we find that $ f(\theta; \phi) $ for different $ \phi $ are fitted well by the mixture of VM and WC defined as follows:
\begin{align}
	\mathrm{MVW}(\theta, \kappa, \rho)
	&= c \mathrm{VM}(\theta, \kappa) + (1-c) \mathrm{WC}(\theta, \rho).
	\label{eq:mvw}
\end{align}
Here, the parameter $ c $ denotes the mixing weight of the two distributions.
We also show the dependence of $ c $ on $ \phi $ for all teams in fig. \ref{fig:Dag-op_op-c_Dag-07}(b).
Remarkably, there exists a point ($ \phi \simeq 0.7 $) at which the value of $ c $ changes drastically.
When $ \phi < 0.7 $, $ f(\theta, \phi) $ follows WC because $ c $ is almost zero, and it approaches VM with increasing $ \phi $ when $ \phi > 0.7 $.
Based on this result, football games are roughly divided into two phases.
Here, we call them {\it disorder phase} ($ \phi < 0.7 $) and {\it order phase} ($ \phi > 0.7 $).
Figure \ref{fig:Dag-op_op-c_Dag-07}(c) shows the angle distributions for disorder phase $ f^{\mathrm{(d)}}(\theta) $ and order phase $ f^{\mathrm{(o)}}(\theta) $.
The parameters for fitting are summarized in table \ref{tb:Dag3}.
They indicate that $ f^{\mathrm{(d)}}(\theta) $ and  $ f^{\mathrm{(o)}}(\theta) $ follow WC and MVW, respectively.
As shown in fig.  \ref{fig:Dag-op_op-c_Dag-07}(c), $ f^{\mathrm{(d)}}(\theta) $ has a fat tail compared with $ f^{\mathrm{(o)}}(\theta) $, namely, the moving direction is more aligned in the order phase.
It is consistent with the result that $ V_{\theta} $ decreases with increasing $ \phi $ (see fig. \ref{fig:op-agsd}).

%


%
\begin{table}[htbp]
  \centering
  \caption{Values of $ c $, $ \kappa $ and $ \rho $ used for fitting of $ f^{\mathrm{(d)}}(\theta) $ and $ f^{\mathrm{(o)}}(\theta) $.}
  \vspace{-0.2cm} 
    \begin{tabular}{ccccc}
    \toprule
 Team                       &  Phase    & $ c $ & $\kappa$ & $ \rho $ \\ \toprule
 \multirow{2}{*}{G10-Home} &  disorder: $ f^{\mathrm{(d)}}(\theta) $ & 0.059 & 21.5     & 0.42   \\ 
                            &  order: $ f^{\mathrm{(o)}}(\theta) $    & 0.45  & 4.62     & 0.73   \\ 
    \bottomrule
    \end{tabular}
  \label{tb:Dag3}
\end{table}

\begin{figure}[htbp]
	\centering
	\includegraphics[width=14cm]{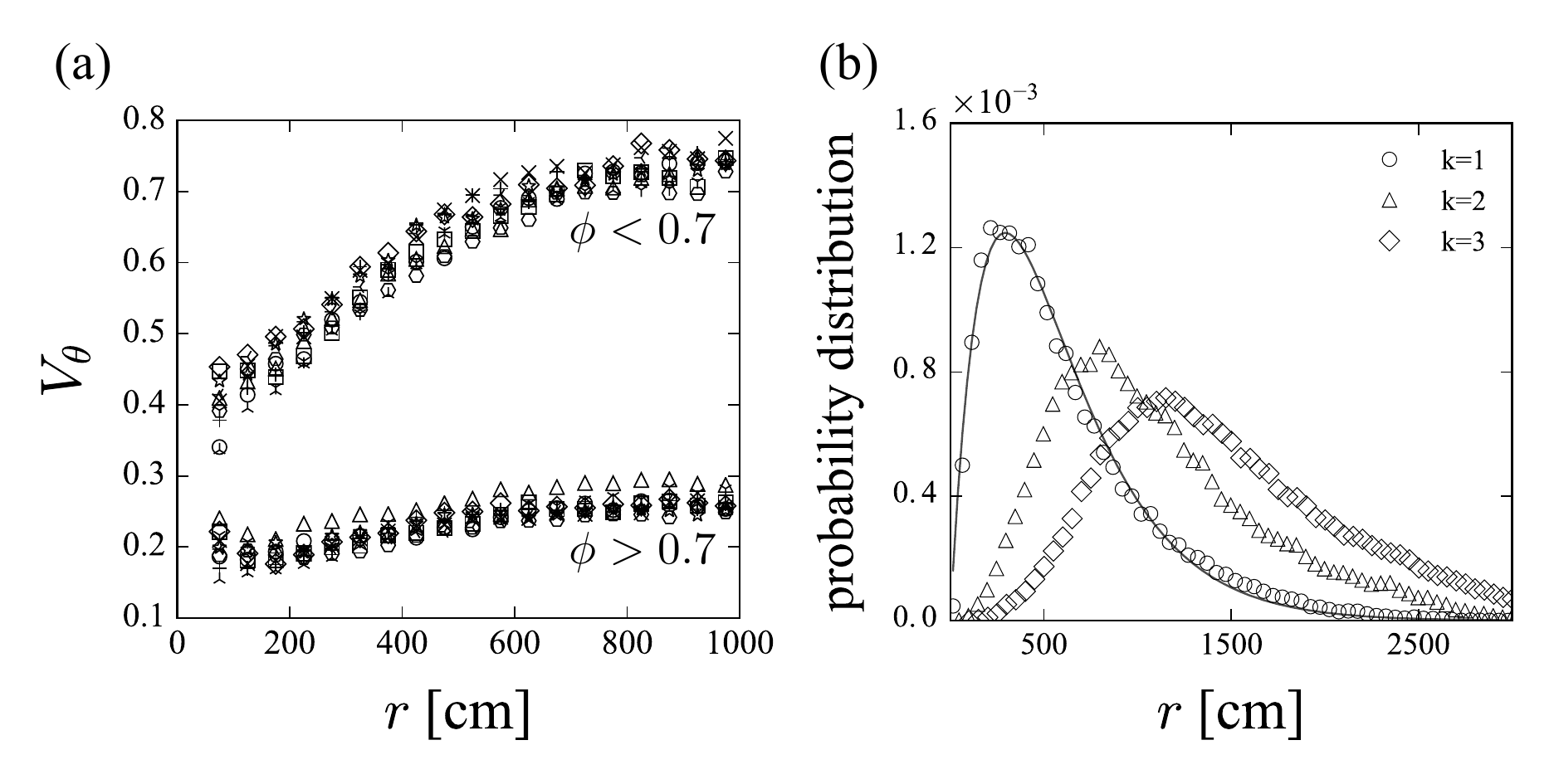}
	\vspace{-0.3cm}
	\caption{(a) Dependence of $ V_{\theta} $ on $ r $ for each phase of all teams. (b) Distribution of interpersonal distance for the $ k $-th nearest pair obtained from the home team in G10. The solid curve shows gamma distribution for fitting.}
	\label{fig:r-agsd_Dr}
\end{figure}

Figure \ref{fig:r-agsd_Dr}(a) shows the dependence of circular variance $ V_{\theta} $ on interpersonal distance $ r $ for each phase of all teams.
$ V_{\theta} $ increases almost linearly with $ r $ when $ r \lesssim 500 $ cm.
On the other hand, it becomes almost constant when $ r \gtrsim 500 $ cm.
Then, $ r \simeq 500 $ cm is a characteristic distance on which the moving direction of interacting players begins to be aligned.
Note that this tendency is significant especially for $ \phi < 0.7 $ since moving directions of all players are aligned regardless of $ r $ when $ \phi > 0.7 $.
Furthermore, we show the distribution of interpersonal distance for the $ k $-th nearest pair in fig. \ref{fig:r-agsd_Dr}(b).
We find that the peak position for $ k=1 $ is less than 500 cm, whereas those for $ k\geq 2 $ are more than 500 cm.
Therefore, the data points with $ r \lesssim 500 $ cm in fig \ref{fig:r-agsd_Dr}(b) are mainly composed of the nearest pair.
This is consistent with the result that the nearest pair exhibits the strong alignment.
We note that the distribution only for the nearest pair ($ k=1 $) follows the gamma distribution (solid curve in fig. \ref{fig:r-agsd_Dr}(b)).
This result is used when we construct a numerical model in the following section.
%

\section{Chase model}

\begin{figure}[htbp]
	\centering
	\includegraphics[width=10cm]{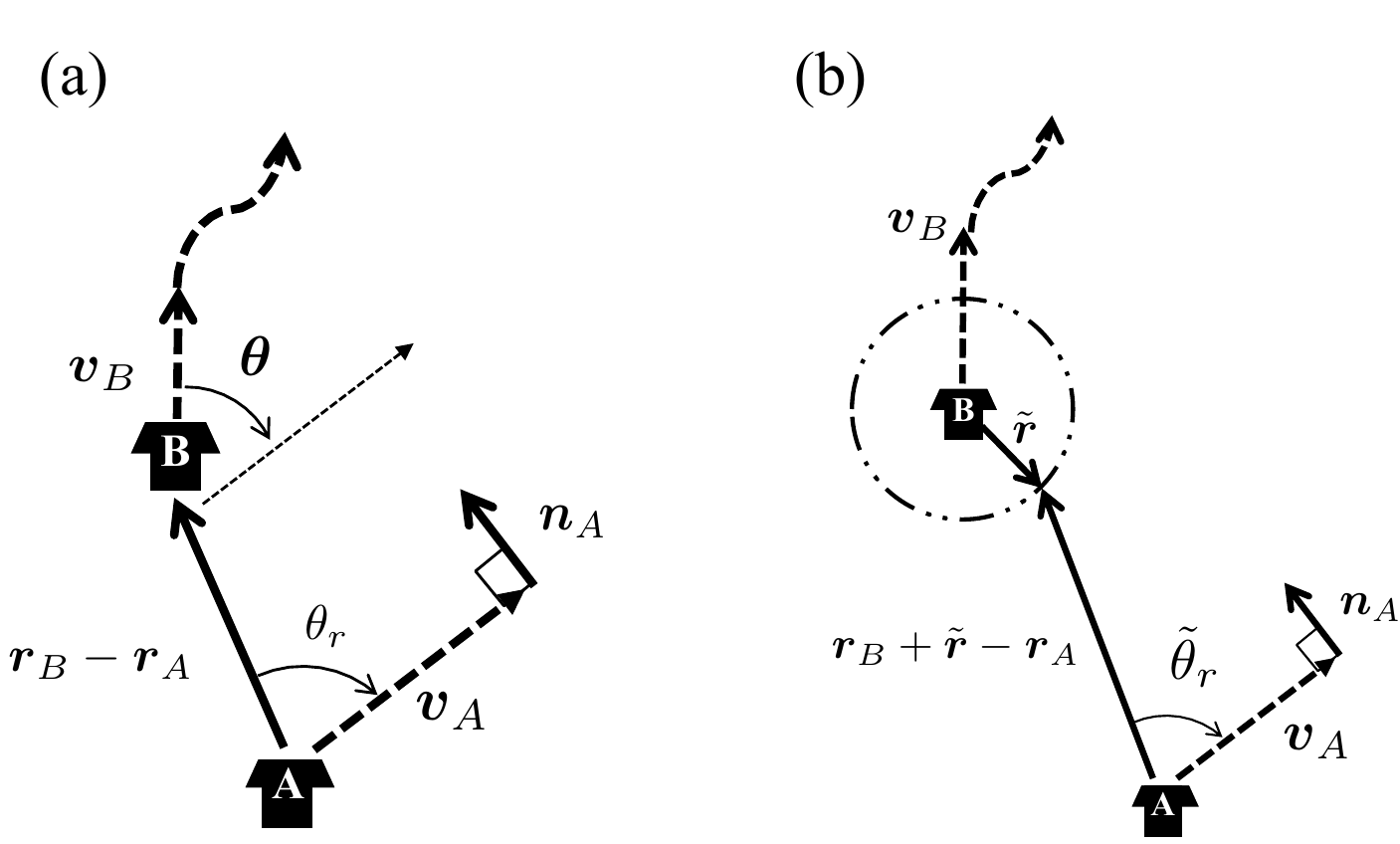}
	\caption{Schematic representation of the two chase models. (a) Original model. (b) Modified model in which $ \tilde{\vector{r}} $ is introduced around the player B.}
	\label{fig:model}
\end{figure}

In this section, we construct a simple model to reproduce the results shown in figs. \ref{fig:Dag-op_op-c_Dag-07}(c) and \ref{fig:r-agsd_Dr}(a).
Now, we consider that the player A chases the player B.
Such chasing can be described as follows:
\begin{subequations}
\begin{align}
	\frac{d \vector{r}_{A}}{d t} &= \vector{v}_{A} \label{eq:eom_A1}\\
	\frac{d \vector{v}_{A}}{d t} &= (v_{0} - |\vector{v}_{A}|)\vector{v}_{A} -  \lambda \sin \theta_{r}\ \vector{n}_{A}, \label{eq:eom_A2}
\end{align} 
\end{subequations}	
\begin{subequations}
\begin{align}
	\frac{d \vector{r}_{B}}{d t} &= \vector{v}_{B} \label{eq:eom_B1}\\
	\frac{d \vector{v}_{B}}{d t} &= (v_{0} - |\vector{v}_{B}|)\vector{v}_{B} +  \vector{\xi}(t).\label{eq:eom_B2}
\end{align} 
\end{subequations}
The first terms of both eqs. \eqref{eq:eom_A2} and \eqref{eq:eom_B2} indicate that each player tends to keep its speed at $ v_{0} $.
In the second term of eq. \eqref{eq:eom_A2}, $ \theta_{r} $ is an angle between $ \vector{v_{A}} $ and $ \vector{r}_{B} - \vector{r}_{A} $, and $ \vector{n}_{A} $ denotes the unit normal vector of $ \vector{v}_{A} $ as shown in fig. \ref{fig:model}(a).
Hence, the player A tends to align $ \vector{v}_{A} $ to $ \vector{r}_{B} - \vector{r}_{A} $, and its tendency generates the chasing behaviour.
The second term of eq. \eqref{eq:eom_B2} is a Gaussian white noise satisfying
\begin{align}
	\langle \vector{\xi}(t) \rangle &= 0, \hspace{0.5cm} \langle \vector{\xi}(t)\cdot \vector{\xi}(t') \rangle = \sigma^{2}\delta(t-t'),
\end{align}
that is, player B moves randomly.

The initial conditions were given as follows in our calculation.
We first set the unit length and time in our model as 10 cm and 0.1 s, respectively.
%
The initial position of player B is set at the origin $ [0, 0] $, and that of player A is set randomly within the circle of radius $ r_{0} = 1000 $ cm whose center is the origin.
Regarding the initial velocity of each player, the speed is set at a constant value $ v_{0} = 120 $ cm/s, and the direction is determined by a uniform random number between $ [-\pi, \pi] $.

\begin{figure}[htbp]
	\centering
	\includegraphics[width=8cm]{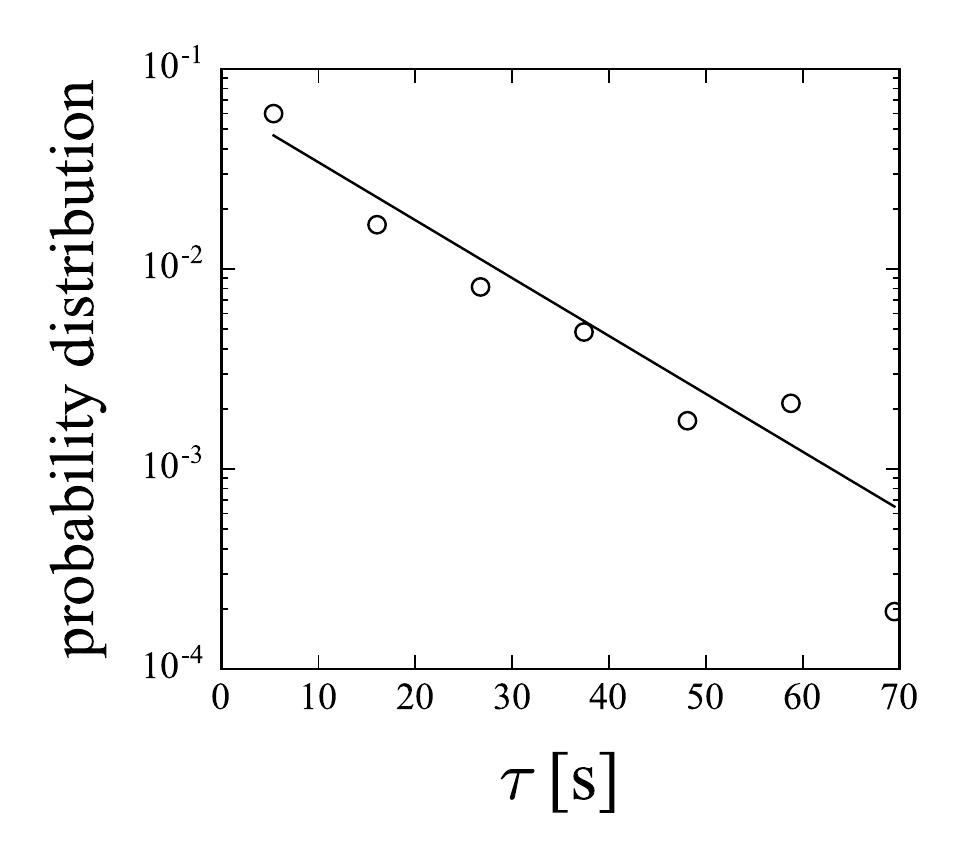}
	\vspace{-0.3cm}
	\caption{Duration time between events in which the ball-possession team changes or the ball is out of the field.}
	\label{fig:dt}
\end{figure}

Here, we add an additional rule in this chase model: reset of chasing.
In real football games, chasing is interrupted by some factors, such as ball-passing, out of play, and so on.
These factors disarrange the alignment of the moving direction of two players.
To take such disarrangement into account, the position and velocity of both players are initialized when $ t=\tau $ [s].
(The initial conditions after reset are the same as explained above.)
$ \tau $ is given as a random variable following the exponential distribution with $ \tau_{0} $ mean.
This assumption is reasonable, because a typical example of duration time between interruptions is characterized by the exponential distribution as shown in fig. \ref{fig:dt}.
We note that the estimation of $ \tau_{0} $ by using real data is difficult because there are many other factors leading to the reset other than the examples in fig. 7, and we have to estimate $ \tau_{0} $ for disorder and order phases separately.


\begin{figure}[htbp]
	\centering
	\includegraphics[width=10cm]{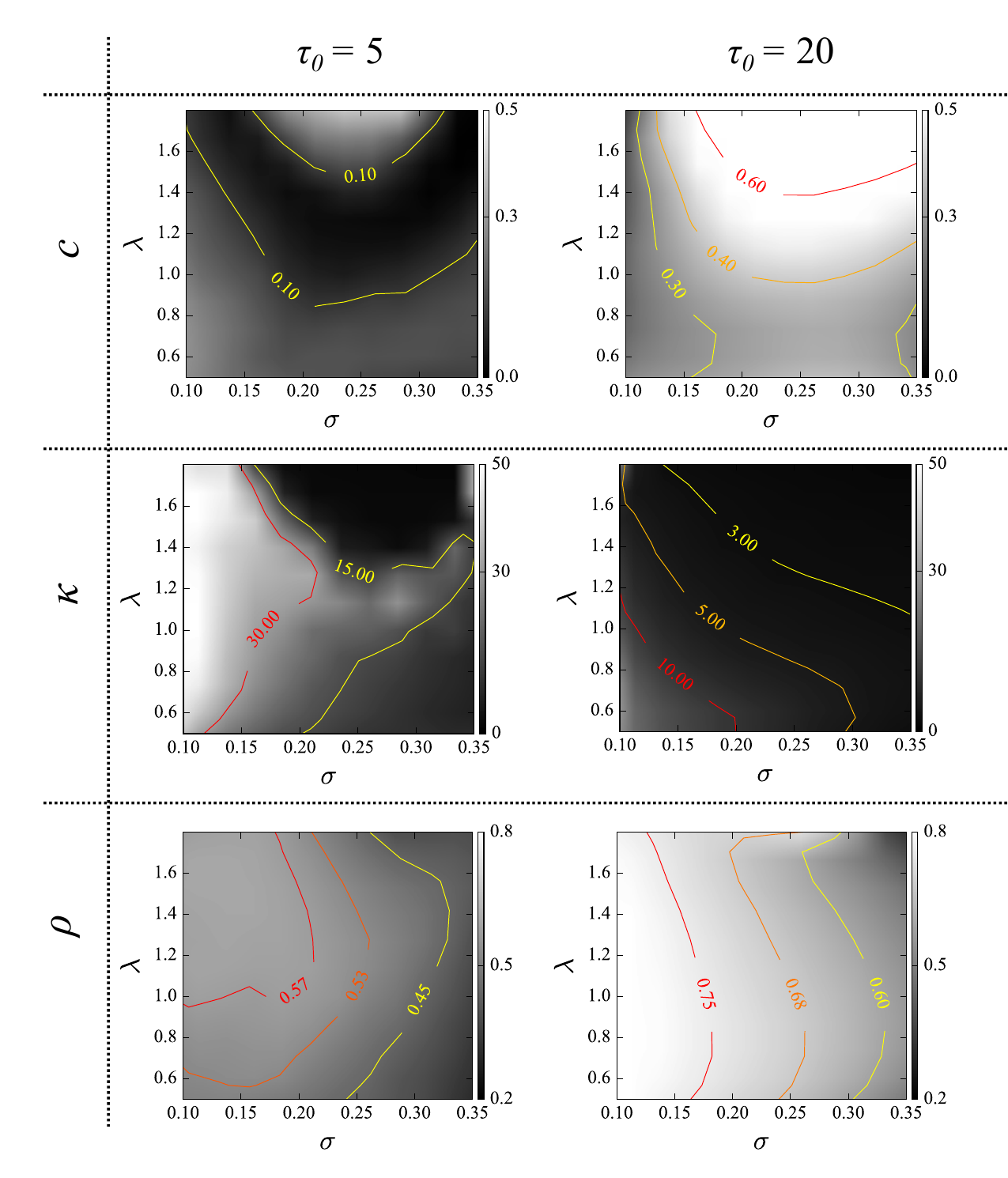}
	\vspace{-0.3cm}
	\caption{Contour plot of $ c, \kappa $ and $ \rho $ as functions of $ \lambda $ and $ \sigma $ for different values of $ \tau_{0} $.}
	\label{fig:ctr}
\end{figure}

\begin{figure}[htbp]
	\centering
	\includegraphics[width=8cm]{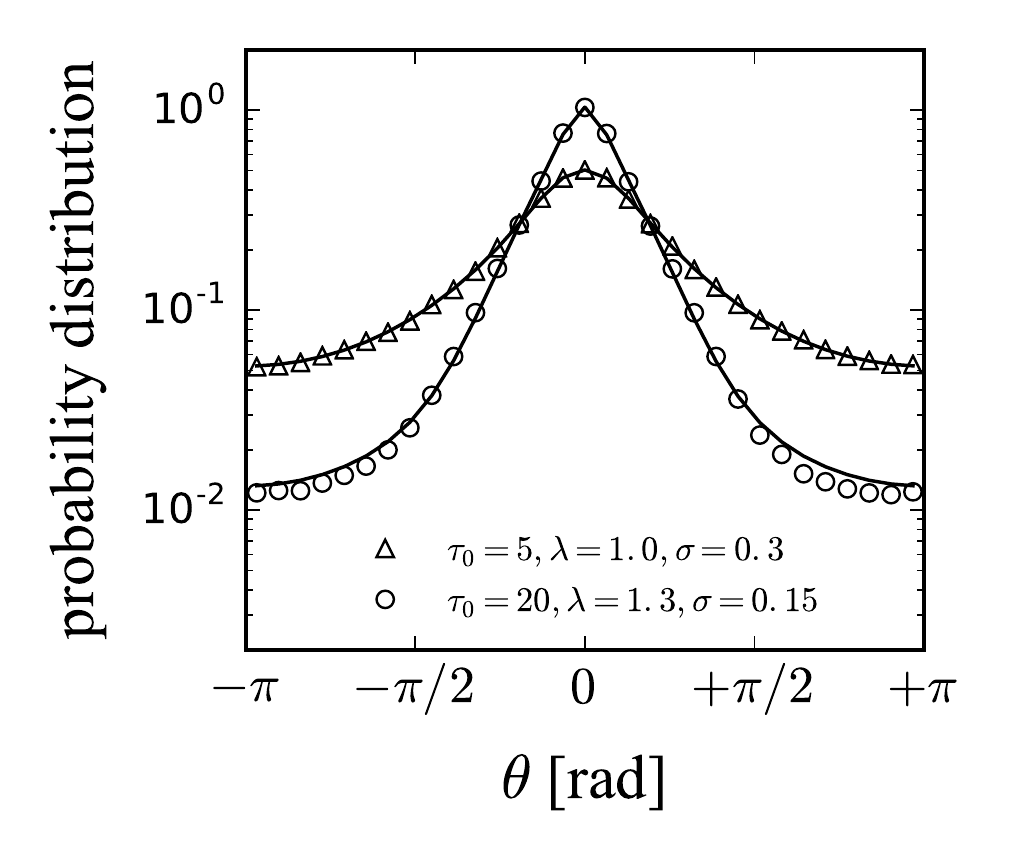}
	\vspace{-0.3cm}
	\caption{Typical examples of the angle distribution obtained from the chase model. The solid curves show the distribution function of eq. \eqref{eq:mvw}, that is MVW.}
	\label{fig:Dag_model}
\end{figure}

In calculation of eqs. \eqref{eq:eom_A1}--\eqref{eq:eom_B2}, we set $ \Delta t = 1 $ and $ 10^{4} $ iterations were executed with the Euler-Maruyama scheme for various parameter sets of $ (\tau_{0}, \lambda, \sigma) $.
For each parameter set, we examine the angle distribution.
It is found that our model reproduces the MVW distribution in wide range of parameter sets.
In fig. \ref{fig:ctr}, we summarize the variations of the fitting parameters $ c, \kappa $ and $ \rho $ of MVW, as functions of controlling parameters $ \lambda $ and $ \sigma $ for different values of $ \tau_{0} $.
We find that there is a tendency that $ c $ increases with increasing $ \tau_{0} $, namely the angle distribution changes from WC to VM with increasing $ \tau_{0} $.
From this figure, we can find the parameter set $ (\tau_{0}, \lambda, \sigma) $ which brings about the similar values of $ c $, $ \kappa $ and $ \rho $ to those from the actual data.
Figure \ref{fig:Dag_model} shows angle distributions when we choose $ (\tau_{0}, \lambda, \sigma) $ as $ (5, 1.0, 0.3) $ and $ (20, 1.3, 0.15) $. 
The solid curves are the result for fitting by using MVW, and the values for fitting by $ (c, \kappa, \rho) $ are shown in table \ref{tb:Dag_model}.
It is found that these values of $ (c, \kappa, \rho) $ are comparable with those for disorder and order phases in table \ref{tb:Dag3} obtained from the real data.
In particular, $ f^{\mathrm{(o)}}(\theta) $ is characterized by a smaller value of $ \sigma $ and larger values of $ \tau_{0} $ and $ \lambda $ compared with $ f^{\mathrm{(d)}}(\theta) $. 
\begin{table}[htbp]
  \centering
  \caption{Values of $ c,\ \kappa $ and $ \rho $ used for fittings of angle distributions shown in fig \ref{fig:Dag_model}.}
    \vspace{-0.2cm}
    \begin{tabular}{ccccccc}
    \toprule
 \multicolumn{3}{l}{Controlling parameters}                      &       &     \multicolumn{3}{l}{Fitting parameters} \\ \cmidrule{1-3}\cmidrule{5-7}
 $\tau_{0}$ & $\lambda$ & $ \sigma $   &   & $ c $ &$ \kappa $ & $ \rho $                                      \\ \toprule
 5  & 1.0  & 0.3  &   & 0.05 & 11.36 & 0.48  \\
 20 & 1.3  & 0.15 &   & 0.39  & 4.51 & 0.76  \\
    \bottomrule
    \end{tabular}
    %
  \label{tb:Dag_model}%
\end{table}%

\section{Discussion}
In the present study, we have shown that WC and MVW appear in $ \phi \lesssim 0.7 $ and $ \phi \gtrsim 0.7 $, respectively.
Then football games are divided into the order and disorder phases.
The difference of angle distributions in order and disorder phases implies that players' interaction is different in each phase.
In football games, the order parameter increases when the long pass or dribble is made, because all players chase the ball all together.
The change of distribution at $ \phi \simeq 0.7 $ seems to be attributed to the ball-chasing behaviour among players.
On the other hand, the interaction between only the nearest pair is dominant in the disorder phase.

We have numerically shown that our model reproduces MVW, and the mixing weight $ c $ mainly depends on the typical duration time $ \tau_{0} $ of chasing.
The reset event randomly disarranges the alignment of moving direction.
Hence, the reset is regarded as an event adding another noise to $ \theta $.
The change of the distribution is attributed to the frequency of adding the  noise: it is more frequent in the disorder phase than in the order phase.

Our proposed model is focused on the directional alignment of two players by the pair interaction between them.
Then, it is difficult to obtain $ \phi $ from our model, since $ \phi $ is defined as the collective property of 20 players in a game.
Nevertheless, our model has a tendency that the directional alignment, which is related to $ \phi $, becomes stronger with decrease of $ \sigma $ and increase of $ \tau_{0} $ and $ \lambda $.

\begin{figure}[htbp]
	\centering
	\includegraphics[width=9cm]{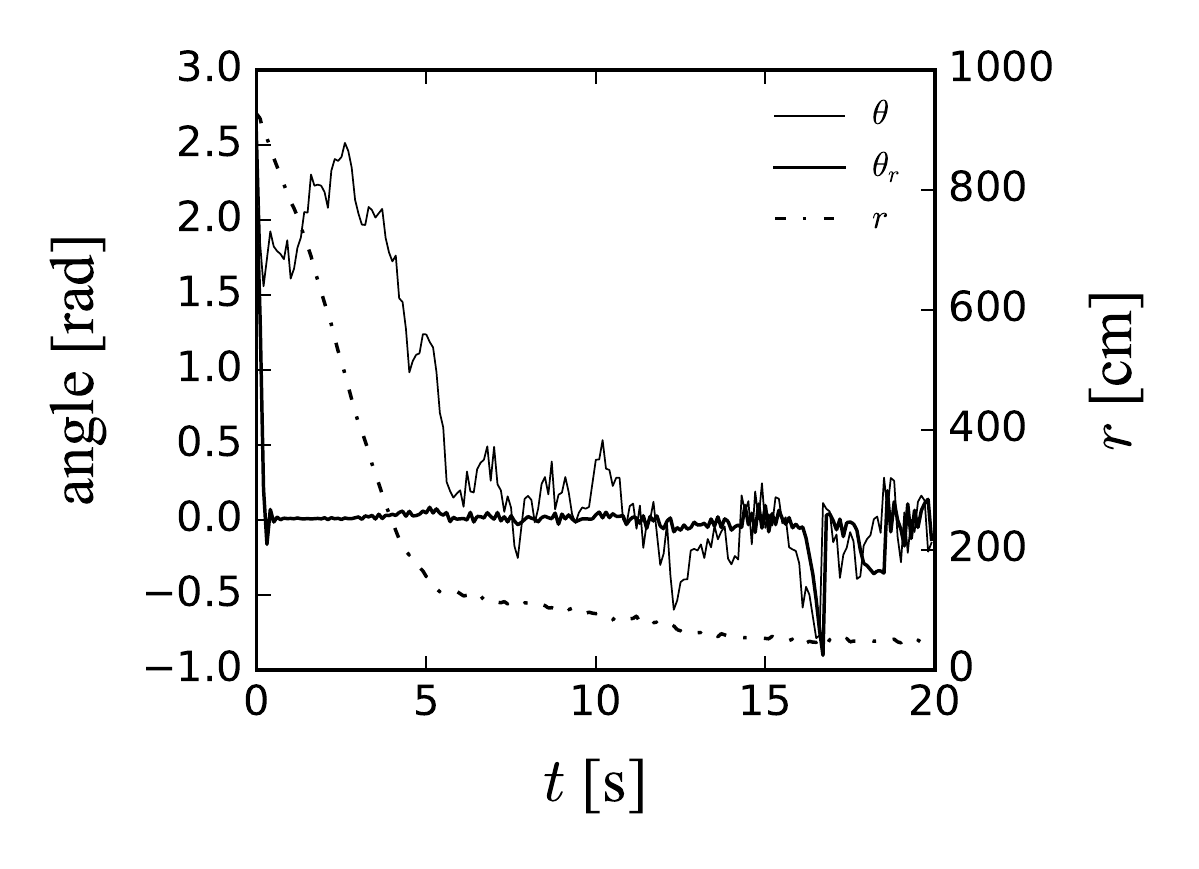}
	\vspace{-0.3cm}
	\caption{Comparison between $ \theta_{r} $ and $ \theta $ when $ \tau=20 $ s and $ \lambda=1.6 $. It is found that $ \theta_{r}\simeq \theta $ is satisfied for small $ r $.}
	\label{fig:ag_agr-t}
\end{figure}

Next, we consider the result that $ f(\theta; \phi) $ approaches VM with increasing $ \phi $ (see fig. \ref{fig:Dag-op_op-c_Dag-07}(c)).
In our model, VM is obtained when the values of $ \tau_{0} $ and $ \lambda $ become large, as shown in fig. \ref{fig:ctr}.
On the other hand, VM is known to be derived from the following Langevin equation
\begin{align}
	\frac{d \theta}{d t}
	&= -\lambda\sin\theta + \xi',
	\label{eq:langevin}
\end{align}
where $ \xi' $ is a Gaussian white noise satisfying $ \langle \xi' \rangle = 0 $ and $ \langle \xi'(t_{1}) \xi'(t_{2}) \rangle = \sigma'^{2}\delta(t_{1}-t_{2}) $.
The corresponding Fokker-Planck equation,
\begin{align}
	\frac{\partial f(\theta, t)}{\partial t}
	&= \frac{\partial}{\partial \theta} \lambda\sin\theta f(\theta, t) + \frac{\sigma'^{2}}{2} \frac{\partial^{2}}{\partial\theta^{2}} f(\theta, t),
\end{align}
has VM as a stationary solution, where $ \kappa = 2\lambda/\sigma'^{2} $.
$ \theta_{r} $ in eq. \eqref{eq:eom_A2} can be approximated as $ \theta $, and then eq. \eqref{eq:eom_A2} and \eqref{eq:eom_B2} generate eq. \eqref{eq:langevin} when $ \tau $ and $ \lambda $ are large and two players are close together.
Actually, fig. \ref{fig:ag_agr-t} shows a typical example ($ \tau=20 $ and $ \lambda=1.6 $).
It is confirmed from this figure that $ \theta_{r}\approx \theta \approx 0 $ when $ t\gtrsim 5  $.
From $ \theta \approx 0 $, $ \vector{v}_{A} $ is almost parallel to $ \vector{v}_{B} $, then the directional alignment becomes strong in this case.
Thus, the dynamics of $ \theta $ in large $ \phi $ can be roughly described by eq. \eqref{eq:langevin}.
In addition, the property that $ f^{(\mathrm{d})}(\theta) $ becomes WC is considered to originate from the uniform noise added at the reset event.

\begin{figure}[htbp]
	\centering
	\includegraphics[width=12cm]{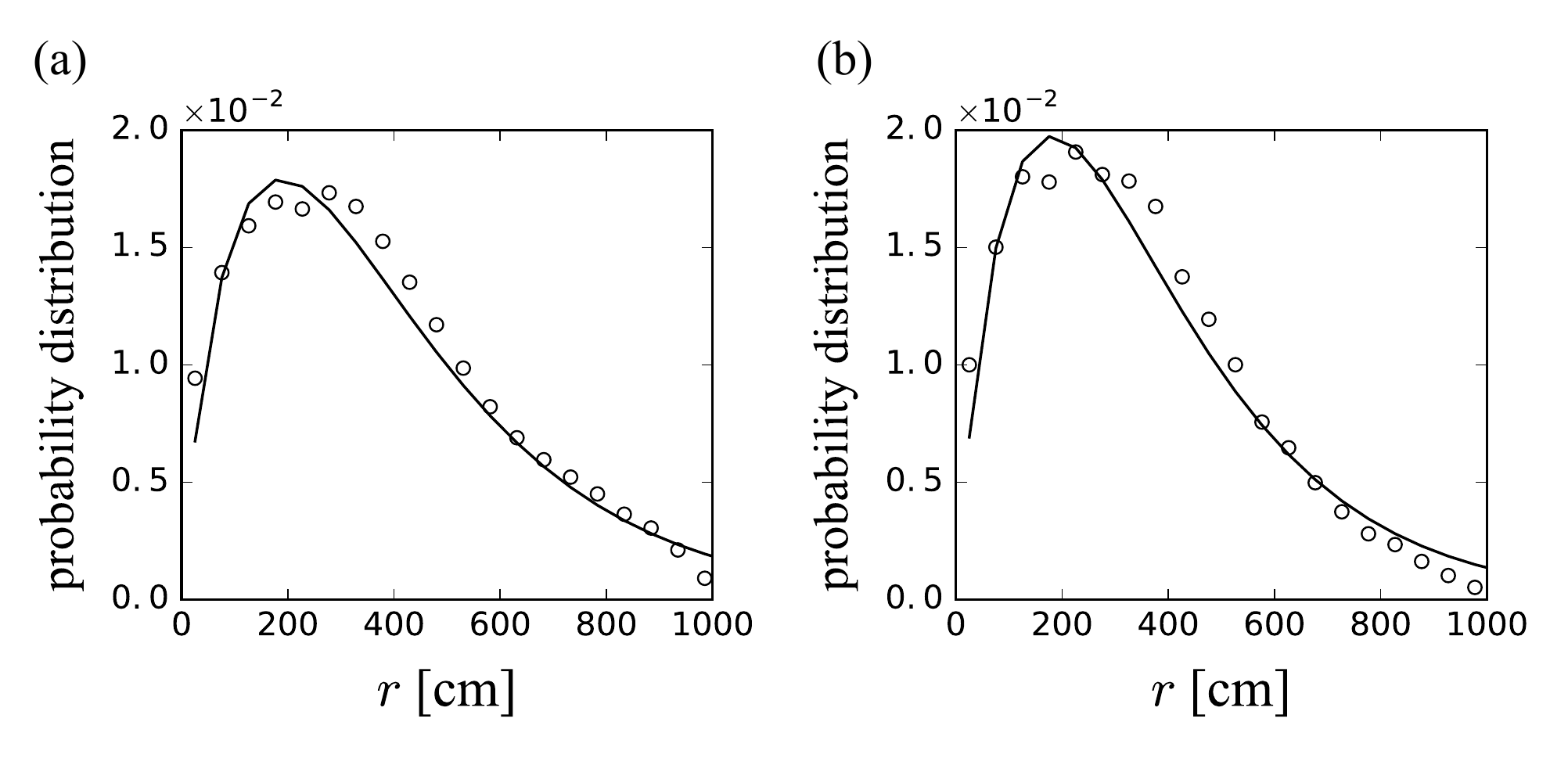}
	\vspace{-0.3cm}
	\caption{Typical examples of the distribution of the interpersonal distance obtained from the modified chase model, where (a) $ (\tau_{0}, \lambda, \sigma) = (5, 1.0, 0.3) $ and (b) $ (\tau_{0}, \lambda, \sigma) = (20, 1.3, 0.15) $. The solid curve in each panel shows the fitting results using the gamma distribution.}
	\label{fig:Dr1_model}
\end{figure}

It is found from fig. \ref{fig:r-agsd_Dr}(a) that there is a characteristic distance, $ r\simeq 500 $ cm, below which $ V_{\theta} $ begins to decrease rapidly as $ r $ decreases.
In order to discuss this result, we consider a more realistic model in which the player A keeps a certain distance during the chasing.
As shown in fig. \ref{fig:model}(b), we introduce $ \tilde{\vector{r}} $ around the player B, and define $ \tilde{\theta}_{r} $ as an angle between $ \vector{v_{A}} $ and $ \vector{r}_{B} + \tilde{\vector{r}} - \vector{r}_{A} $.
We assume that the direction of $ \tilde{\vector{r}} $ is given by a uniform random number between $ [-\pi, \pi] $, and that the magnitude $ |\tilde{\vector{r}}| $ is given by a normal random number whose mean and standard deviation are 250 cm and 200 cm, respectively.
Figure \ref{fig:Dr1_model} shows the distribution of interpersonal distance obtained from the numerical calculation results of the modified model.
It is found that these distributions are fitted well by gamma distributions
This result is consistent with that shown in fig. \ref{fig:r-agsd_Dr}(b).

%

\begin{figure}[htbp]
	\centering
	\includegraphics[width=12cm]{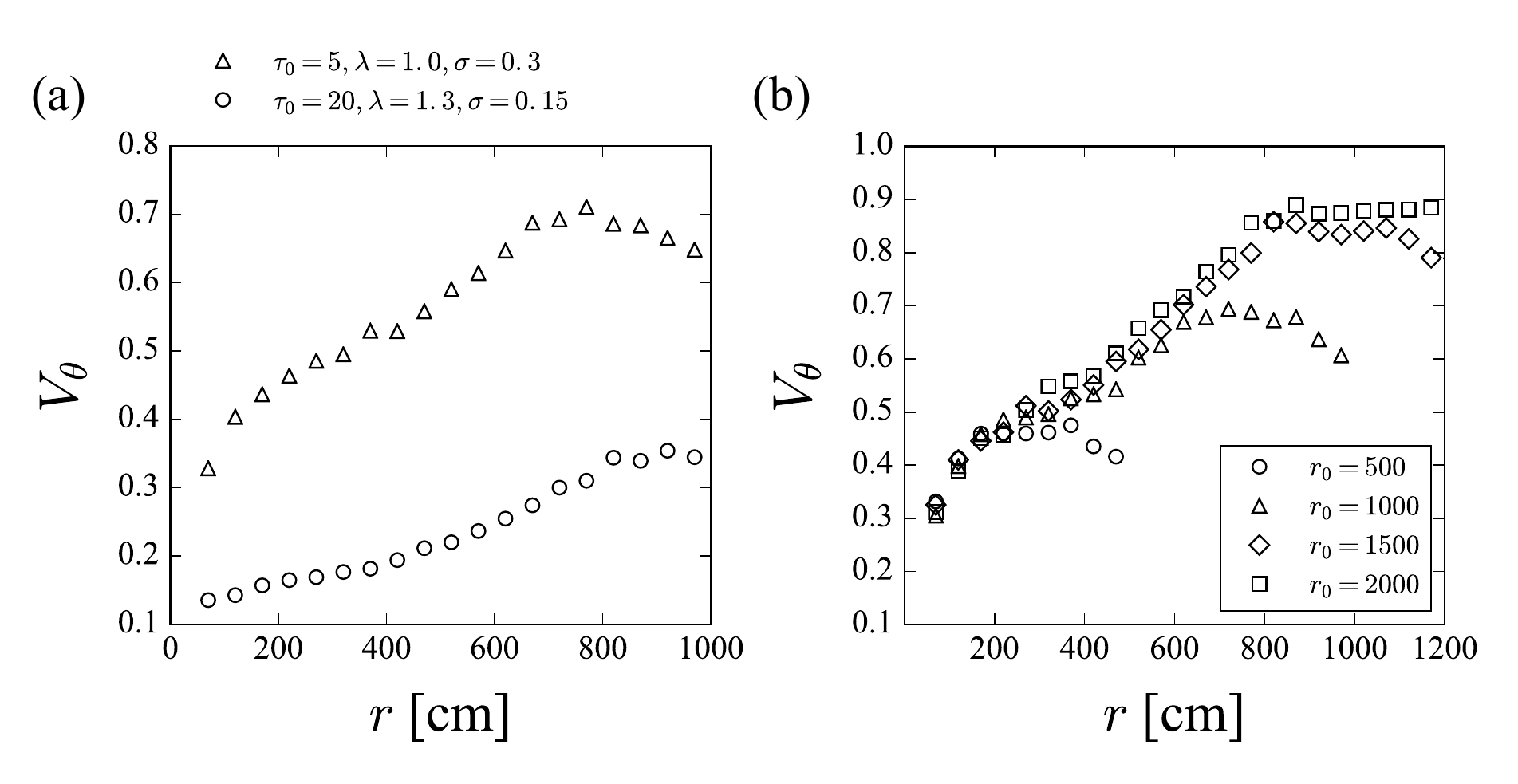}
	\vspace{-0.3cm}
	\caption{(a) Dependence of $ V_{\theta} $ on $ r $ for $ (\tau_{0}, \lambda, \sigma) = (5, 1.0, 0.3) $ and $ (20, 1.3, 0.15) $. (b) Dependence of $ V_{\theta} $ on $ r $ for $ (\tau_{0}, \lambda, \sigma) = (5, 1.0, 0.3) $ for different values of $ r_{0} $.}
	\label{fig:r-cv_model}
\end{figure}

It is confirmed that this modification to the original model (eqs. \eqref{eq:eom_A1}--\eqref{eq:eom_B2}) does not have an significant effect on the angle distributions．
Further, as shown in fig. \ref{fig:r-cv_model}(a), we obtain the dependence of $ V_{\theta} $ on $ r $ for the two parameter sets, $ (\tau_{0}, \lambda, \sigma) = (5, 1.0, 0.3) $ and (20, 1.3, 0.15).
In particular, this figure shows the linear dependence of $ V_{\theta} $ when $ r\lesssim 500 $ cm and $ V_{\theta} $ of the order phase is smaller than that of the disorder phase.
These points are in good agreement with the results shown in fig. \ref{fig:r-agsd_Dr}(a).
These results are reproduced if the player A starts chasing within the circle of radius $ r_{0} = 1000 $ cm.
Then, the chasing distance of players in football games can be estimated as $ r_{0} \simeq 1000 $ cm.
Note that we have confirmed that $ r_{0} \simeq 1000 $ cm is the best to reproduce fig. \ref{fig:r-agsd_Dr}(a) in the range of $ 500 \leq r_{0} \leq 2000 $ as shown in fig. \ref{fig:r-cv_model}(b).
$ r_{0} \simeq 1000 $ cm seems to be consistent with the ordinary football strategy that each player marks opponent players within a certain distance.

We expect that our analysis and modelling can be applied to other sports including basketball and hockey, or systems other than competitive sports, because chasing behaviour is widely observed.
Moreover, the property ``chases and escapes'' is a classical problem in mathematics and game theory  \cite{Nahin, Isaacs}.
In physics, this property has attracted our attention on collective motion of self-propelled particles \cite{Kamimura2010, Nishi2012, Ohira2015}.
We believe that our findings give a new insight into these research fields.

%

\section{Conclusion}
In the present paper, we have investigated the statistical properties for motion of interacting players in football games.
First, we have analysed the player tracking data by focusing on the angle $ \theta $ between two velocity vectors for the pair of one player and the $ k $-th nearest opponent player.
We have examined the circular variance $ V_{\theta} $ and angle distribution as functions of the order parameter $ \phi $ and interpersonal distance $ r $.
It is found that the nearest pair exhibits strong alignment of their moving directions.
In particular, it is revealed that $ r \simeq 500 $ cm is a characteristic distance on which the moving direction of interacting players begins to be aligned.
We have also found that WC and MVW appear at $ \phi \lesssim 0.7 $ and $ \phi \gtrsim 0.7 $, respectively.
Thus, football games can be divided into the order and disorder phases.
Next, we have constructed the chase models.
The point of these models is the reset of chasing.
We have numerically shown that our model can reproduce the statistical properties of player interactions for the nearest pair, including angle distribution in each phase, distribution of interpersonal distance, and dependence of $ V_{\theta} $ on $ r $.
These numerical results indicate that there is another characteristic distance $ r\simeq 1000 $ cm below which a player begins to chase an opponent player.
We conclude that our models embody the essential features of interactions between players in competition.

\section*{Acknowledgements}
The authors are very grateful to DataStadium Inc., Japan for providing the player tracking data. 
The authors thank Drs. K. Yamamoto and H. Nishimori for the fruitful discussion.
This work was partially supported by unit for ``Multi-scale Analysis,
Modelling and Simulation'', Top Global University Project of Waseda
University, and the Data Centric Science Research Commons Project of the Research Organization of Information and Systems, Japan.

\bibliography{./reference} 
\end{document}